\documentclass[twoside]{article}
\usepackage{amsmath}
\usepackage{amsfonts}
\usepackage{amssymb}
\usepackage{amscd}
\usepackage{color}

\def\sumint{\sum\mspace{-25mu}\int}

\begin{document}
\thispagestyle{plain}

\begin{center}
{\Large \bf \strut
Relativistic few-body physics
\strut}\\
\vspace{10mm}
{\large \bf 
Wayne Polyzou$^{abc}$
}
\end{center}

\noindent{
\small $^a$\it The University of Iowa, Department of Physics and Astronomy,
Iowa City, IA }\\ 
\noindent{
\small $^b$\it This work supported by the US Department of Energy
under contract No. DE-FG02-86ER40286}\\
\noindent{
\small $^c$\it For the Proceedings of the 22nd European Few-Body 
Conference in Kracow Poland}    

\markboth{
W. N. Polyzou}
{Relativistic few-body physics} 

\begin{abstract}
 I discuss different formulations of the relativistic 
few-body problem with an emphasis on how they are related.  I first
discuss the implications of some of the differences with
non-relativistic quantum mechanics.  Then I point out that the
principle of special relativity in quantum mechanics implies that the
quantum theory has a Poincar\'e symmetry, which is realized by a
unitary representation of the Poincar\'e group.  This representation
can always be decomposed into direct integrals of irreducible
representations and the different formulations differ only in how
these irreducible representations are realized.  I discuss how these
representations appear in different formulations of relativistic
quantum mechanics and discuss some applications in each of these
frameworks.
\\[\baselineskip] 
\end{abstract}

\section{Introduction}
In these proceedings I discuss different formulations of the
relativistic few-body problem.  The goal is to understand the relation
between different formulations of relativistic quantum mechanics that
are used in applications.  The fundamental assumption underlying all
of these formulations is the principle of special relativity in
quantum mechanics, which implies that in a box isolated from the rest
of the universe it is impossible to use experimental measurements to
distinguish different inertial coordinate systems.

In quantum mechanics the experimentally measurable quantities are
probabilities, expectation values and ensemble averages.  These are
preserved if the states and observables identified with different
inertial coordinate systems are related by unitary transformations.
Wigner \cite{Ref1} showed that this condition is necessary and
sufficient for the invariance of these observables with respect to
changes of the inertial coordinate system.  What he showed is that
relativistic invariance in the sense described in the previous
paragraph is equivalent to the existence of a unitary representation,
$U(\Lambda ,a)$, of the Poincar\'e group (Lorentz $+$ space-time
translations).  This is common to all relativistic quantum mechanical
formulations of the few-body problem.  In addition, unitary
representations of the Poincar\'e group can always be decomposed into
a direct integrals of irreducible representations.  This decomposition
is the relativistic analog of diagonalizing the Hamiltonian in
non-relativistic quantum mechanics.  States in each irreducible
subspace transform among themselves via well-defined transformations
under the Poincar\'e group.  All of the relevant observables can be
computed directly from the irreducible eigenstates.

These irreducible representations are common to all formulations of
relativistic few-body quantum mechanics.  The equivalence of different
models is determined by the equivalence of decompositions into
irreducible representations.  Different formulations of relativistic
quantum mechanics differ how the Hilbert spaces are represented, which
can vary from being abstract to concrete.   

We discuss how irreducible representations of the Poincar\'e group
appear in theories that emphasize Lorentz covariance, Poincar\'e
covariance, Euclidean covariance, as well as 
$2j+1$ vs. $4j+2$ component spinors.

\section{Poincar\'e group}
\label{sec2}

The Poincar\'e group has ten infinitesimal generators that satisfy
commutation relations characteristic of the Poincar\'e Lie algebra.
They are the Hamiltonian, $H$, the linear momentum, $\mathbf{P}$, the
angular momentum $\mathbf{J}$, and the rotationless Lorentz boost
generators $\mathbf{K}$.  Three of the commutators have the
Hamiltonian on the right hand side of the commutation relations
\begin{equation}
[P^m,K^n] = i \delta_{mn} H .
\label{eq:1}
\end{equation}
Since interactions appear on the right hand side of these commutation 
relations, between
three and ten generators must include interactions \cite{Ref2}.  
This is needed to ensure the consistency of the many possible ways 
of representing time evolution.  

\section{Cluster properties}
\label{sec3}
 
The discussion of performing experiments in a box that is isolated
from the rest of the universe requires that the unitary representation
of the Poincar\'e group factors into a tensor product of a
representation involving the degrees of freedom in the box and another
representation involving the remaining degrees of freedom, when the
degrees of freedom in the box are isolated from the rest of the
universe.

I refer to this property as cluster separability of the unitary representation
of the Poincar\'e group.  It means for states describing asymptotically 
separated subsystems, $s_1$ and $s_2$, that    
\begin{equation}
U(\Lambda ,a) \to U_{s_1} (\Lambda ,a) \otimes  U_{s_2} (\Lambda ,a).
\label{eq:1b}
\end{equation}
It implies that each generator has a cluster expansion and the
generators for each subsystem also must satisfy the Poincar\'e
commutation relations.  Because some interactions involve particles in
overlapping subsystems, the additional invariance requirements cannot
be satisfied without introducing a special class of many-body
interactions that are distinguished by being functions of the subsystem
interactions.  Weinberg \cite{Ref3} suggested that avoiding these
difficulties is an important motivation for using local field theory.
In spite of the difficulties, Sokolov \cite{Ref4} has solved the
problem of constructing the many-body interactions that are needed to restore
cluster properties.

If a model satisfies cluster properties in one frame, such as the rest
frame, then the invariance of the $S$-matrix means the $S$-matrix
will satisfy cluster properties in all frames, even when the unitary
representation of the Poincar\'e group does not.  However, when this
system is embedded in a larger system, it will not always be
in the rest frame of the larger system.  In this case the many-body 
interactions need to be included to restore cluster properties in 
the rest frame of the larger system.

What happens in a three-body model that satisfies cluster properties
in the three-body rest frame, but not at the level of the Poincar\'e
generators is that if one particle is asymptotically separated from an
interacting pair of particles in frames other than the rest frame, the
interaction between the pair of particles vanishes.  Violations of 
cluster properties have observable consequences, however these appear
to be small for nuclear reactions \cite{Ref5}\cite{Ref6}

\section{Position in relativistic quantum theory}
\label{sec4}

Another property of relativistic quantum mechanics is the absence
of a consistent position operator.  The standard explanation for this 
is that the energy needed to localize a particle in a sufficiently small
spatial region is enough to create pairs of particles and 
antiparticles.  However, one does not need antiparticles understand
the problem. 
If one assumes that a state representing a particle at the 
origin at time $t=0$ is Lorentz invariant, then properties of the 
irreducible representations imply that the overlap of states with 
spacelike separated position do not identically vanish; instead they
the fall off exponentially with a scale determined by the particle's
Compton wavelength.  Formally
\[
\langle m, \mathbf{p} \vert x \rangle = 
\langle m, \mathbf{p} \vert U(I,x) \vert 0 \rangle = 
\]
\[
e^{i p \cdot x} \langle m, \mathbf{p} \vert 0 \rangle =
e^{i p \cdot x} \langle m, \mathbf{p} \vert 
U(\Lambda (p),0) \vert 0 \rangle =
\]
\begin{equation}
e^{i p \cdot x} \sqrt{{m \over \omega_m (\mathbf{p})}} 
\langle m, \mathbf{0}  
\vert 0 \rangle . 
\label{eq:2}
\end{equation}
This implies 
\begin{equation}
\langle \mathbf{x},0 \vert \mathbf{y},0 \rangle = 
\int e^{i \mathbf{p} \cdot (\mathbf{x}-\mathbf{y})}
{c \over \omega_m(\mathbf{p})} d\mathbf{p} =
2 \pi i c {K_1 (m \vert \mathbf{x} -\mathbf{y} \vert)
\over  m \vert \mathbf{x} -\mathbf{y} \vert} .
\label{eq:3}
\end{equation}
For large $x$ the Bessel function $K_1(x)$ decays exponentially
in $x$.

One thing that follows from this observation is that the notation of
retardation, which is a field observable, it is not an observable for
relativistic particles, since they cannot be precisely localized. 
   
\section{Spin}
\label{sec5}

The treatment of spin seems to distinguish irreducible representations
of the Poincar\'e group from Lorentz covariant theories.
Specifically, for spin 1/2 particles, four-component spinors are used in
Lorentz covariant theories while spin 1/2 irreducible representations
of the Poincar\'e group involve only two-component spinors.  In an
experiment measurements of the spin projection can take on only two
values.

The relation between these two equivalent treatments of 
spin can be understood by considering 
the transformation properties of a Poincar\'e irreducible 
basis state 
\[
U(\Lambda,0) \vert (m,j) \mathbf{p}, \mu \rangle =
\]
\begin{equation}
\sum \vert (m,j) \pmb{\Lambda}{p}, \mu' \rangle
\sqrt{{ \omega_m (\pmb{\Lambda}p) \over \omega_m (\pmb{p})}}
\underbrace{D^j_{\mu' \mu}[B^{-1}
(\pmb{\Lambda} p/m) \Lambda B(\mathbf{p}/m)]}_{\mbox{\bf Wigner rotation}}. 
\label{eq:4}
\end{equation}
Because $D^j_{\mu' \mu}[R] $ is also a $2j+1$ dimensional 
representation of $SL(2,\mathbb{C})$ it can be used 
to decompose the Wigner rotation into a product of finite dimensional
representations of Lorentz transformations.  
Lorentz covariant states can be defined by
\begin{equation}
\vert {p},j, \sigma \rangle_c := \sum
\vert (m,j) \mathbf{p}, \mu \rangle \sqrt{\omega_m (\mathbf{p})}
D^j_{\mu \sigma}[ B^{-1}(\mathbf{p}/m)].
\label{eq:5}
\end{equation}
With this definition eq. (\ref{eq:4}) becomes 
a Lorentz covariant transformation of the covariant states:
\begin{equation}
U (\Lambda ,0) \vert p,j, \sigma \rangle_c  = \sum
\vert \Lambda p ,j, \sigma' \rangle_c
D^j_{\sigma'\sigma}[\Lambda] . 
\label{eq:6}
\end{equation}
The scalar product of two covariant wave-functions, 
\begin{equation}
\Phi (\mathbf{p}, \sigma) :=
_c\langle {p},j, \sigma \vert \Phi \rangle ,
\label{eq:8}
\end{equation}
is
\begin{equation}
\langle \Psi \vert \Phi \rangle =
\sumint \Psi^* (\mathbf{p}, \sigma)  
{m d \mathbf{p}  \over \omega_{m}(\mathbf{p}) }
D^j_{\sigma \sigma'} [ B(\mathbf{p}/m)
B^{\dagger} (\mathbf{p}/m)]
\Phi (\mathbf{p}, \sigma').
\label{eq:7}
\end{equation}
The new feature is that the Lorentz covariant inner product has a
momentum and spin-dependent kernel, which maintains the unitarity of the
representation.  The problem is that while for $SU(2)$,
$U=\sigma_2U^{*}\sigma_2$, there is no similarity transformation relating
representations of $SL(2,C)$ to the complex conjugate representations.
Technically this is not needed, because the Wigner functions of the 
boosts and Lorentz transformation
are eventually combined to get a Wigner rotation, where this difference 
disappears,
however the problem is that space 
reflection relates these inequivalent representations.  Since in the 
representation (\ref{eq:7}) boosts appear in the kernel of the 
inner product, the kernel and hence the representation of the 
Hilbert space will change under space reflection.  It is possible 
to get a Hilbert space representation of space reflection by 
making the replacement
\begin{equation}
D^j_{\sigma \sigma'} [ B(\mathbf{p}/m)
B^{\dagger} (\mathbf{p}/m)] 
\to 
D^j_{\sigma_1 \sigma_1'} [ B(\mathbf{p}/m)
B^{\dagger} (\mathbf{p}/m)] \oplus 
D^j_{\sigma_2 \sigma_2'} [ B(-\mathbf{p}/m)
B^{\dagger} (-\mathbf{p}/m)]. 
\label{eq:9}
\end{equation}

For spin $1/2$ this means replacing two-component spinors by
four-component spinors.  When the Lorentz transformations are combined
with the boosts, the Wigner rotations for
both representations become identical, recovering the original
Poincar\'e covariant result.

\section{Non-relativistic limits}
\label{sec6}

In the strong coupling region of QCD there is not enough mathematical
control of the theory for precision calculations of scattering
reactions.  As a consequence, realistic models of the strong
interaction are ultimately fit to nucleon-nucleon scattering data.
This is certainly the case for the so-called high-precision nuclear
potentials.  These fits are normally performed by first Lorentz
transforming the cross section data from the laboratory frame to the center
of momentum frame; and then interactions are adjusted so the cross
section in the non-relativistic model reproduces this data with a
$\chi^2$ per degree of freedom near $1$.  Relativistic interactions
can be computed the same way, replacing the non-relativistic model by a
relativistic one.  The fits are performed by expressing the cross
section in terms of transition matrix elements that are functions the
rest-frame momentum of one of the particles.

When this is done, both models will give the same phase shifts as a
function of the relative momentum that appears in the scattering
amplitude.  When this procedure is used there are no relativistic
effects in two-nucleon observables.  Relativistic effects first appear
in the three-body problem, where cluster properties dictate how to
embed moving two-body subsystems in the three-particle Hilbert space.

This does not mean that the non-relativistic limit of a 
relativistic nucleon-nucleon model is small.  This 
distinction is important.

This situation in strong interaction physics is analogous to the one
that one would have if the Coulomb potential was replaced by a
non-relativistic phenomenological potential that reproduced all fine
and hyperfine structure in the Hydrogen atom.  In that case 
there would be no two-body relativistic corrections.  The difference
with the strong interactions is that there is no method comparable to 
QED for constructing realistic nucleon-nucleon from QCD. 

The important conclusion is that it makes no sense to compute
relativistic corrections to a non-relativistic model when the
interaction in both models is fit to the same scattering data.
 
\section{Hilbert space representations}
\label{sec7}

The most familiar relativistic quantum theory is relativistic
quantum field theory.  It is instructive relate the description of
a free particle in Minkowski quantum field theory, Euclidean 
quantum field theory and Poincar\'e invariant quantum mechanics.
Different relativistic few-body models are related to one of these
representations.
 
In a free quantum field theory a one-particle state is 
constructed by smearing the field with a test function of 
four spacetime variables and applying the resulting operator
to the vacuum.  The Hilbert space scalar product is 
\[
(f,f) = \langle 0 \vert \phi(f^*) \phi(f) \vert 0 \rangle = 
\int f(x)^*
\langle 0 \vert 
\phi(x) \phi(y) \vert 0 \rangle
f(y) d^4x d^4y = 
\]
\[
\int f(x)^* {1 \over (2 \pi)^3}
\int d^4p \theta (p^0) \delta (p^2 + m^2) e^{i p \cdot (x-y)}
f(y) d^4x d^4y =
\]
\begin{equation}
\int \tilde{f}^*(\omega_m(\mathbf{p}), \mathbf{p})
{d \mathbf{p} \over 2 \omega_m (\mathbf{p}) }
\tilde{f}(\omega_m(\mathbf{p}), \mathbf{p})
\label{eq:10}
\end{equation}
where 
\begin{equation}
\tilde{f}(\omega_m(\mathbf{p}), \mathbf{p}) = {1 \over (2 \pi)^{3/2}}
\int d\mathbf{x} e^{i \mathbf{p}\cdot \mathbf{x} -i \omega_m(\mathbf{p})}
g(x_0,\mathbf{x}) .
\end{equation}
The corresponding Hilbert space inner product in Euclidean quantum 
field theory is 
\[ 
(g,g) =
\int g(-x_0 ,\mathbf{x} )^* S(x-y) g(y_0,\mathbf{y}) d^4x_e d^4y_e
\]
\[
\int g(-x_0,\mathbf{x} )^* {d^4p \over (2\pi)^4}{e^{i p_e \cdot (x-y) }
\over p^2 +m^2} 
g(y_0,\mathbf{y}) d^4x_e d^4y_e
\]
\begin{equation}
= \int {\tilde{g}^*(x_0,\mathbf{p})} 
{e^{- \omega_m (\mathbf{p})x_0}} 
{dx_0 dy_0 d\mathbf{p} \over 2 \omega_{m}(\mathbf{p})} 
e^{- \omega_m (\mathbf{p})y_0}
\tilde{g}(y_0,\mathbf{p})
\label{eq:11}
\end{equation}
where 
\begin{equation}
\tilde{g}(x_0,\mathbf{p}) = {1 \over (2 \pi)^{3/2}}
\int d\mathbf{x} e^{-i \mathbf{p} \cdot \mathbf{x}}
g(x_0,\mathbf{x}). 
\label{eq:12}
\end{equation}
Both expressions lead to integrals of functions of 
the three momentum integrated with a Lorentz invariant 
measure, which is the inner product associated with 
a mass $m$ irreducible representation of the Poincar\'e group.

Comparing these expressions implies the following 
relations between these equivalent representations of a 
free spinless particle of mass $m$:
\[
\Psi (\mathbf{p})  =  
 {1 \over (2 \pi)^{3/2}}\langle \mathbf{p} \vert \phi(f) \vert 0 \rangle 
=\int f(\mathbf{x},t) e^{-i \omega_m (\mathbf{p})t - i \mathbf{p} \cdot \mathbf{x}}
{d^4x \over \sqrt{2\omega_{m}(\mathbf{p})}} 
\]
\begin{equation}
= {1 \over (2 \pi)^{3/2}}\int g(\mathbf{x},x_0) 
e^{- \omega_m (\mathbf{p}) x_0 - i \mathbf{p} \cdot \mathbf{x}}
{d^4x \over \sqrt{2\omega_{m}(\mathbf{p})}} .
\label{eq:13}
\end{equation}

These relations determine different realizations of the unitary representations
of the Poincar\'e group. In the Minkowski 
quantum field theory case the manifestly covariant transformation 
\begin{equation}
f(x) \to f_{\Lambda ,a}(x) = f(\Lambda^{-1}( x-a))
\label{eq:14}
\end{equation}
is unitary provided the vacuum expectation value of the fields is
invariant
\begin{equation}
W(x-y) = W(\Lambda (x - y)) = \langle 0 \vert \phi(x) \phi(y) \vert 0 \rangle.  
\label{eq:15}
\end{equation}
The relations (\ref{eq:13}) and the representation (\ref{eq:14})
lead to explicit expressions for the Poincar\'e generators 
for both Euclidean field theory as well as Poincar\'e irreducible 
quantum theory.  The Hamiltonian and boost generators in these two case 
are found to be 
\begin{equation}
H g(x) = {\partial g \over \partial \tau} (x) \qquad
K_i g(x) = \tau {\partial g \over \partial x_i} (x)-
x_i {\partial g \over \partial \tau} (x)
\label{eq:16}
\end{equation}
and
\begin{equation}
H \Psi (\mathbf{p}) = \sqrt{\mathbf{p}^2 + m^2} \Psi (\mathbf{p})
\qquad
K_i \Psi (\mathbf{p}) = {i \over 2}\{ H ,
{\partial  \over \partial p_i}  \} 
\Psi (\mathbf{p}).
\label{eq:17}
\end{equation}
The generators for rotations and translations are the same in all
three representations. 

The main features to notice are that in both the Euclidean and
Minkowski field theory cases the formula for the covariant inner
product has a non-trivial kernel.  These considerations generalize to
interacting theories and the generalization of (\ref{eq:13}) to
interacting theories implies precise relations between the different
formulations of relativistic quantum mechanics.  The relations between
the wave functions imply relations between the corresponding unitary
representations of the Poincar\'e group and their infinitesimal
generators

\section{Realizations}
\label{sec7}

In what follows I discuss different ways of constructing relativistic 
few-body models.  I focus on the structure of the model Hilbert space and 
representations of the Poincar\'e group.

\subsection{Covariant constraint dynamics}
\label{sec:8}

In this framework the dynamics is given by solving a set of coupled
constraint equations 
\begin{equation}
C_1 \psi = (p_1^2 + m_1^2 + V_1)\psi =  0
\label{eq:18}
\end{equation}
\begin{equation}
C_n \psi = (p_n^2 + m_n^2 + V_n)\psi = 0
\label{eq:19}
\end{equation}
where integrability requires that constraints commute on solutions
(first class constraints)  
\begin{equation}
[C_i, C_j] \psi = 0 .
\label{eq:20}
\end{equation}
A typical realization is 
\begin{equation}
[p_1^2 - p_2^2 , V ] \psi = 0 .
\label{eq:21}
\end{equation}
Because the constraints are covariant and satisfy the 
first class condition they can be used to make a model of a
Wightman function, $W(x-y)$  
of the form
\begin{equation}
W = \prod_i \delta (C_i).  
\label{eq:22}
\end{equation}
In this case the inner product is 
\[
\langle g \vert f \rangle =  
\]
\begin{equation}
\int g^*(x_1, x_2)
\langle x_1, x_2 \vert \delta (C_1)\delta (C_2) \vert y_2, y_1 \rangle f (y_1,y_2)
d^8x d^8y
\label{eq:23}
\end{equation}
and the transformation 
\begin{equation}
f (y_1,y_2) \to f (\Lambda y_1+a ,\Lambda y_2 + a )
\label{eq:24}
\end{equation}
defines a unitary representation for the Poincar\'e group acting on the 
two-particle Hilbert space.  This method has been used in a number of 
applications involving both constituent quarks \cite{Ref7}\cite{Ref8} 
and charged 
particles \cite{Ref9}\cite{Ref10}. 
 
\subsection{Direct interaction quantum mechanics}
\label{sec:9}

This formulation of relativistic quantum mechanics is formulated on 
direct sums of tensor products of single-particle irreducible representation
spaces of the Poincar\'e group.   Interactions are added to sums of 
single-particle generators in a manner that preserves the Poincar\'e commutation 
relations.
\begin{equation}
{\cal H} = \oplus (\otimes {\cal H}_{jm})
\label{eq:25}
\end{equation}
\begin{equation}
H = \sum_i H_i + \sum_{ij} H_{ij} + \sum_{ijk} H_{ijk} + \cdots 
\label{eq:26}
\end{equation}
\begin{equation}
\mathbf{P} = 
\sum_i \mathbf{P}_i + \sum_{ij} \mathbf{P}_{ij} + \sum_{ijk} 
\mathbf{P}_{ijk} + \cdots 
\label{eq:27}
\end{equation}
\begin{equation}
\mathbf{J} = 
\sum_i \mathbf{J}_i + \sum_{ij} \mathbf{J}_{ij} + \sum_{ijk} 
\mathbf{J}_{ijk} + \cdots 
\label{eq:28}
\end{equation}
\begin{equation}
\mathbf{K} = 
\sum_i \mathbf{K}_i + \sum_{ij} \mathbf{K}_{ij} + \sum_{ijk} 
\mathbf{K}_{ijk} + \cdots 
\label{eq:29}
\end{equation}
This method has the advantage that techniques used to solve non-relativistic
few-body models can be directly applied to relativistic few-body problems. 
This method has been used extensively in applications; including
constituent quark models \cite{Ref11}\cite{Ref12}\cite{Ref13}, 
nuclear reactions \cite{Ref14}\cite{Ref15}\cite{Ref16}, and electromagnetic 
probes of hadronic systems \cite{Ref17}

\subsection{Manifestly covariant methods}
\label{sec:10}

These methods involve using equations of quantum
field theory that relate different time-ordered Green functions.  The
simplest of these equations is the Bethe-Salpeter equation.  Formally
the Bethe-Salpeter Kernel is defined as the difference of the inverse
of a product of two-point Green functions and the inverse of a
four-point Green function:
\begin{equation}
K = - (G^{-1})_c := G_0^{-1}- G^{-1}.  
\label{eq:30}
\end{equation}
This is equivalent to the Bethe-Salpeter
equation for the four point function, $G$,
\begin{equation}
G = G_0 + G_0 K G .
\label{eq:31}
\end{equation}
In applications $G_0$ and $K$ must be modeled.  The advantage of these
covariant methods is that both spectra and scattering observables can
be calculated directly from $G$, without any direct use of the
underlying relativistic quantum theory; however the formulas for
calculating observables require complete sets of Poincar\'e
irreducible eigenstates between products of fields.

Translational invariance implies the Fourier transforms have the from
\begin{equation}
\tilde{G}(P) = \tilde{G}_{0}(P) + \tilde{G}_{0}(P) \tilde{K}(P) 
\tilde{G}(P) .
\label{eq:32}
\end{equation}
The existence of a two-body irreducible intermediate states implies 
that $\tilde{G}(P)$ has a pole a $P^2=m^2$:
\begin{equation}
\tilde{G}(P) = -2 \pi i {\chi (P) \bar{\chi} (P) \over P^0-E} + \cdots  
\label{eq:33}
\end{equation}
where the Bethe-Salpeter wave function $\chi (P)$ is related  to the underlying 
irreducible representation of the Poincar\'e group by 
\begin{equation}
\chi (P) = \int \langle (j,m)\mathbf{P},\mu \vert T(\phi (X+{1 \over 2}x) 
\phi(X-{1\over 2}x)) 
\vert 0 \rangle
{e^{i P \cdot X} \over (2 \pi)^2 } d^4X .  
\end{equation}

The residue at the pole satisfies the homogeneous  Bethe-Salpeter equation
\begin{equation}
\chi (P) = \tilde{G}_0(P)\tilde{K}(P) \chi (P) 
\qquad P^2 = - M^2 .
\label{eq:34}
\end{equation}
The normalization of the residue is related to the normalization of 
$G$ by
\begin{equation}
1= {i \over 4\pi P^0 } \chi(P) {\partial \tilde{G}^{-1}(P) \over \partial P^0} 
\bar{\chi}(P) .
\label{eq:35}
\end{equation}
While the Bethe-Salpeter wave-functions $\bar{\chi}(P)$ do not 
represent probability amplitudes,
given the normalization condition (\ref{eq:35}) they can be used to compute 
matrix elements of observables in Poincar\'e irreducible eigenstates
\cite{Ref18}.  

These methods are most useful in theories like QED where $K$ and $G_0$  
can be reliably approximated using perturbation theory.  Applications of 
these methods are discussed in the contributions by Bakker \cite{Ref19}, 
Karmanov \cite{Ref20}, and Salm\'e \cite{Ref21} in this volume.

\subsection{Quasipotential methods}
\label{sec:11}

The manifestly covariant methods have the computational 
disadvantage that the integral equations involve both space and 
time variables (or four momentum variables).  Quasipotential 
methods reformulate the covariant Bethe-Salpeter equation 
into an equivalent pair of equations.  The first step is to 
write the product of two-point functions as the sum of a constrained 
function with the same poles, and a difference, 
\begin{equation}
\tilde{G}_0(P) = \tilde{g}_0 (P) + \tilde{\Delta} (P) .
\label{eq:36}
\end{equation}
This leads to an equation for a quasipotential, $\tilde{U}(P)$, 
that replaces the Bethe-Salpeter kernel, 
\begin{equation}
\tilde{U} (P) = \tilde{K}(P) + \tilde{K}(P) \tilde{\Delta} (P) \tilde{U}(P) 
\label{eq:37}
\end{equation}
and a constrained four-point function that satisfies a quasipotential 
equation with one less continuous integration variable than the 
Bethe-Salpeter equation
\begin{equation}
\tilde{g}(P) = \tilde{g}_0(P) + \tilde{g}_0(P) \tilde{U}(P) \tilde{g}(P).
\label{eq:38}
\end{equation}
The solution to this equation has the same poles as the Bethe-Salpeter equation.
The residues are related to the residues of the Bethe-Salpeter  equations 
\begin{equation}
\tilde{K}(P) \chi (P)  =  \tilde{U} (P) \xi (P) 
\label{eq:39}
\end{equation}
and they satisfy a normalization condition 
\begin{equation}
1= {i \over 4\pi P^0} \xi(P) {\partial g^{-1}(P) \over \partial P^0} \bar{\xi}(P).
\label{eq:40}
\end{equation}
In most applications the quasipotental, $\tilde{U}(P)$,  is modeled 
directly, rather than by solving (\ref{eq:37}) with a model 
kernel $\tilde{K}(P)$.  

The residues can be used in the same way that the $\chi (P)$ are used
to calculate observables in the Bethe Salpeter equation.

Example of covariant quasipotential equations are the Gross equation
\cite{Ref22}\cite{Ref23}.

\subsection{Euclidean methods}
\label{sec:12}

Euclidean relativistic quantum mechanics is a formulation of relativistic 
quantum mechanics that models the Hilbert space inner product 
used in the Euclidean formulation of quantum field theory by 
modeling the Euclidean Green functions that appear in the 
kernel of the Hilbert space scalar product
\begin{equation}
\langle f \vert g \rangle 
 = \sum_{mn} \int f^*_m(Rx)S_{m+n} (x,y )  g_n (y) d^{4m}x d^{4n}y ,
\label{eq:41}
\end{equation}
where $S_{m+n} (x,y )$ is an $m+n$ point Euclidean Green function,
$R$ is Euclidean time reflection, and the functions $f$ and $g$
are non-zero only when the relative times are positive.
The collection of Green functions is called reflection positive  
if $\langle f \vert f \rangle \geq 0$.  Poincar\'e generators are
constructed from Euclidean generators as follows; Euclidean generators
that anticommute with $R$ are multiplied by $i$ and generators 
that commute with $R$ remain unchanged. The resulting operators
are Hermitian with respect to the inner product (\ref{eq:41}) and satisfy 
the Poincar\'e commutation relations.   The dynamics can be 
expressed in terms of $e^{-\beta H}$,
which is represented by a translation of all Euclidean times by 
$\beta >0$.  This operator can be used like the Hamiltonian to 
compute bound and scattering observables.

This method is closely related to Euclidean version of the 
Schwinger-Dyson equations.  The advantage of the quantum mechanical
interpretation is that all calculations can be performed in the
Euclidean framework without analytic continuation.  
Model calculations have been used to demonstrate 
that $e^{-\beta H}$ can be used to calculate GeV-scale 
scattering observables \cite{Ref24} using the relation  
\begin{equation}
\langle \psi \vert \Omega^{\dagger}_+ \Omega_- \vert \phi \rangle 
\qquad \Omega_{\pm} = 
\lim_{n \to \pm \infty} e^{-ine^{-\beta H}}\Pi e^{+ine^{-\beta H_0}}
\label{eq:42}
\end{equation}
where $\Pi$ is a mapping from an asymptotic channel space to the 
Hilbert space.

\section{Summary}

In this contribution I demonstrated the unifying role played by
irreducible representations of the Poincar\'e group in essentially all
formulations of relativistic few-body quantum theory.  These
representations arise as a result of the requirement that all quantum
mechanical observables have the same value in all inertial coordinate
systems, which requires that the dynamics is given by a unitary
representation of the Poinar\'e group.  The irreducible representation
are the elementary building blocks of these dynamical representations. 

The author would like to acknowledge and
  thank numerous collaborators who have significantly contributed the
  physics discussed in these proceedings. This includes Fritz Coester,
  Charlotte Elster, Walter Gl\"ockle, Jacek Golak, Hirouyki Kamada,
  Brad Keister, Phil Kopp, Ting Lin, Roman Skib\'inski, Henryk
  Wita{\l}a and Yunfei Huang.


\begin{thebibliography}{3}
%
%
\bibitem{Ref1} E. P. Wigner (1939), Ann. Math. 40,140.
\bibitem{Ref2} P. A. M. Dirac (1949), Rev. Mod. Phys. 21,392.
\bibitem{Ref3} S. Weinberg (1995), The Quantum Theory of Fields, Vol 1, Cambridge University Press, Ch. 4.
\bibitem{Ref4}S. N. Sokolov (1977), Dokl. Akad, Nauk SSSR, 233, 575.
\bibitem{Ref5} B. Keister and W. N. Polyzou (2012), Phys. Rev. 
C 86,014002.
\bibitem{Ref6}F. Coester and W. N. Polyzou (1982), Phys. Rev. 
D 26,1348.
\bibitem{Ref7}H. W. Crater, R. Becker, C. Y. Wong, and P. Van Alstine (1992), 
Phys. Rev.  D 46, 5117. 
\bibitem{Ref8}H. W.  Crater and C.-Y. Wong (2012), Phys. Rev.,
D 85, 116005.
\bibitem{Ref9}H. W.  Crater and Peter Van Alstine, (1988) Phys. Rev. D 37, 1982.
\bibitem{Ref10}H. W. Crater, James Schiermeyer, (2010) Phys. Rev.  D 82,094020.
\bibitem{Ref11} R.F. Wagenbrunn, S. Boffi, W. Klink, W. Plessas, and M. Radici (2001), 
Phys. Lett. B511, 33.
\bibitem{Ref12} Wolfgang Schweiger (these proceedings)
\bibitem{Ref13} S. Capstik and B. D. Keister Capstik,
 Phys.Rev. D 51 3598.
\bibitem{Ref14}T. Lin, Ch. Elster, W. N. Polyzou, W. Gl\"ockle (2008), Physics
Letters,  B660, 345.
\bibitem{Ref15}H. Wita{\l}a, J. Golak, R. Skib\'inski, W. Gl\"ockle, H.
Kamada, and W. N. Polyzou, (2011), Phys. Rev. C 83, 044001.
\bibitem{Ref16}  Emanuele Pace, Giovanni Salme, Sergio Scopetta, 
Alessio Del Dotto, Matteo Rinaldi (2013), Few Body Syst. 54, 1079-1082.
\bibitem{Ref17} V.A. Karmanov, A.V. Smirnov, (1994) 
Nuc. Phys., A575, 520-548
\bibitem{Ref18} K. Huang and A. Weldon (1975), Phys. Rev. 
D 11,257.
\bibitem{Ref19} B. Bakker (these proceedings).
\bibitem{Ref20} V.A. Karmanov, J. Carbonell
(these proceedings).
\bibitem{Ref21} G. Salm\'e (these proceedings). 
\bibitem{Ref22} Franz Gross and Alfred Stadler, (2008), 
Phys. Rev.  C 78, 014005.
\bibitem{Ref23} Elmar P. Biernat, Franz Gross, Teresa Pe˜na, Alfred Stadler (these proceedings).
\bibitem{Ref24} P. Kopp and W. N. Polyzou (2012), Phys. Rev. D 85,016004.

\end{thebibliography}


\end{document}